\documentclass[twocolumn,prd,preprintnumbers,amsmath,amssymb]{revtex4}
\usepackage{graphicx}
\usepackage{times}
\usepackage{slashed}
\usepackage{amsmath}
\usepackage{epsfig}
\def\beq{\begin{equation}}
\def\ee{\end{equation}}
\def\eeq{\end{equation}}

\def\bfig{\begin{figure}}
\def\efig{\end{figure}}
\def\bea{\begin{eqnarray}}
\def\beann{\begin{eqnarray*}}
\def\eea{\end{eqnarray}}
\def\eeann{\end{eqnarray*}}

\def\3p0{$^{3}P_{0}$}

\begin{document}

\title{On the possibility of a 130 GeV gamma-ray line from annihilating singlet fermionic dark matter}


\author{T. H. Franarin, C. A. Z. Vasconcellos, D. Hadjimichef}
\affiliation{
Instituto de F\'{\i}sica, Universidade Federal do Rio
  Grande do  Sul\\
 Av. Bento Gon\c{c}alves, 9500\\ Porto Alegre, Rio Grande do Sul,
CEP 91501-970, Brazil
}



\begin{abstract}
There is evidence for a spectral line at $E_\gamma\approx 130$ GeV in the Fermi-LAT data that can be explained as dark mater particles annihilating into photons. We review a well known dark matter model that consists in a singlet Dirac fermion and a singlet scalar. The scalar implements  spontaneous symmetry breaking in the dark sector, and is responsible for the communication between dark matter and Standard Model particles through a coupling to the Higgs. These interactions are supressed by the mixing between the scalar and the Higgs. Therefore, the singlet fermionic dark matter is naturally a weakly interacting massive particle (WIMP) and can explain the observed relic density. We show that this model cannot produce the signal identified in the Fermi-LAT data. Thus, we propose a modification in the model by  introducing a new scalar multiplet that carries electric charge and couples to the singlet scalar. It enhances the annihilation into two photons and succeeds in producing the observed signal. We also discuss the resulting increase of the branching ratio of the $h\rightarrow\gamma\gamma$ process, which is consistent with measurements from the CMS experiments.

\end{abstract}

\maketitle

\section{Introduction}

The problem of dark matter (DM) in the universe is one of the major mysteries of cosmology, astrophysics and particle physics. According to the Planck satellite observations, and based on the standard model of cosmology, the total mass-energy of the known universe contains 4.8\% ordinary matter, 26.1\% dark matter and 69.1\% dark energy (Planck Collaboration 2013). 

The first indication of the existence of dark matter (DM) came from Jeans (1922) analysis of the vertical motions of stars near the plane of the Milky Way. Zwicky (1933) measured radial velocities of galaxies in the Coma Cluster and concluded that it must contain a large amount of dark matter to hold galaxies together. Many other observations have indicated the existence of dark matter, including the rotation curves of galaxies, gravitational lensing of background objects by galaxy clusters, and the fluctuations in the cosmic microwave background. 

So far, the most direct observational evidence for dark matter is the Bullet Cluster, where a collision between two galaxy clusters have caused a separation of dark matter and ordinary matter. X-ray observations show that much of the hot gas has been slowed down by ram pressure and settled near the point of impact. However, weak gravitational lensing observations show that much of the mass is distributed outside of the central region. Because dark matter does not interact by electromagnetic forces, the DM halos of the two clusters would have passed through each other without slowing down substantially (Clowe et al. 2004). 

All the found evidences for DM come from its gravitational effects. So far, searches for explicit signals of DM particles have been giving negative results. Indirect detection aim to find signals of the annihilations of DM particles in the fluxes of cosmic rays. Usually, the searches consist in looking for channels and ranges of energy where it is possible to overcome the background from ordinary astrophysical processes.

Weniger (2012) have found tentative evidence for a line spectral feature 
at $E_\gamma\approx 130$ GeV in the Fermi-LAT data. It can be explained by $130$ GeV dark matter annihilating into two photons with a cross section approximately 24 times smaller than that needed for the thermal relic density. Afterwards, the signal was independently confirmed (Tempel \& Hektor \& Raidal 2012; Su \& Finkbeiner 2012a; Hektor \& Raidal \& Tempel 2012b). In some analysis, another line at $E\approx 110$ GeV is observed (Su \& Finkbeiner 2012b). The double peak at 110 GeV and 130 GeV was also observed in the gamma-ray excess from eighteen nearby galaxy clusters (Hektor \& Raidal \& Tempel 2012a). This feature is a generic prediction of DM annihilation, corresponding to $\gamma\gamma$ and $\gamma Z$ final states. The fact that the signals from the Galactic Center (GC) and from galaxy clusters give excesses that precisely coincide suggests that this is not a statistical fluctuation.

Profumo \& Linden (2012) have suggested that the signal has an astrophysical origin associated with the Fermi bubble regions, but Tempel et al. (2012) claims that the strongest emission is coming from close to the Galactic Center and not from the Fermi bubbles regions. Aharonian, Khangulyan \& Malyshev (2012) proposed that ultrarelativistic pulsar winds could be the source of the gamma-ray line. The possibility that the gamma-ray line is a fake instrumental effect is disfavored, as discussed in (Finkbeiner, Su \& Weniger 2012) and (Hektor et al. 2012b). 

The signal was confirmed by the Fermi-LAT collaboration. Using reprocessed data, the feature shifted to 133 GeV and lost statistical signiﬁcance (Fermi-LAT Collaboration 2013). More data are needed to clarify its origin. We adopt the annihilating DM hypothesis and analyze if a fermionic dark matter could account for it.

This work is organized as follows.  In Sec. 2, we present the WIMP paradigm. In Sec. 3, we report the gamma-ray line detected in the Fermi-LAT data. In Sec. 4, we review a fermionic dark matter model. In Sec. 4 we study the constraint imposed by the observed DM relic density. In Sec. 6, we calculate the annihilation into two photons. In Sec. 7, we show the results and discussions, introducing a modification in the model. Finally, we give our conclusion in Sec. 8.

\section{WIMPs}

Several generic classes of DM candidates have been suggested and the most widely accepted is weakly interacting massive particles (WIMPs). The interest in this approach comes from the fact that WIMPs in thermal equilibrium in the early universe naturally give the right abundance to be cold dark matter. Furthermore, these same interactions make the detection of WIMPs possible.

It is assumed that in the early universe WIMPs were produced and annihilated in particle-antiparticle collisions, such as
\begin{equation}
\chi\bar{\chi}\longleftrightarrow e^+e^-, \mu^+\mu^-, q\bar{q}, W^+W^-, ZZ, HH,...
\end{equation}

At temperatures much higher than the WIMP mass, the colliding particle-antiparticle pair in the thermal plasma had enough energy to create WIMP pairs efficiently. The annihilation reactions of WIMPs into Standard Model (SM) particles were initially in equilibrium with the production processes.

As the universe expanded, the temperature of the plasma dropped below the WIMP mass. While maintaining equilibrium, the number of WIMPs produced decreased exponentially as $e^{-m_\chi/T}$. Meanwhile, the expansion of the universe decreased the number density of particles $n$, and with it the production and annihilation rate. When the WIMP annihilation rate became smaller than the expansion rate of the universe, the interactions of WIMPs froze out and their number density in a comoving volume remained approximately constant. 

The relic density of WIMPs is approximately
\begin{equation}
\Omega h^2\approx 0.1\times\frac{3\times 10^{-26}\text{cm}^3\text{s}^{-1}}{\langle\sigma_{\text{ann}}v\rangle_F},
\label{miracle}
\end{equation}
where $\langle\sigma_{\text{ann}}v\rangle_F$ is the thermally averaged annihilation cross section at freeze-out. The most precise measurement of the relic density has been obtained from the Planck satellite data as $\Omega_{\chi}h^{2}=0.1199\pm 0.0027$ (Ade et al. 2013), where $h$ is the Hubble constant in units of $100\,  \text{km}\, \text{s}^{-1}\, \text{Mpc}^{-1}$. From Eq. (\ref{miracle}), we see that weak scale particles give the right relic density. This fact is known as the WIMP miracle.

\section{Gamma-ray line}

The gamma-ray flux produced by the annihilation of self-conjugated DM particles $\chi$ (e.g. Majorana fermions) into two photons inside the Galactic DM halo is
\begin{equation}
\frac{dJ_\gamma}{dEd\Omega}(\theta)=\frac{\langle\sigma v\rangle_{\chi\chi\rightarrow\gamma\gamma}}{8\pi m^2_\chi}2\delta(E-E_\gamma)\int_{l.o.s.}ds\rho^2(r),
\label{fluxo}
\end{equation}
where $\theta$ is the angle between the direction of the line of sight and the axis connecting the Earth to the GC. Here, $m_\chi$ is the DM mass, $\langle\sigma v\rangle_{\chi\chi\rightarrow\gamma\gamma}$ the thermally averaged cross section for $\chi\chi\rightarrow\gamma\gamma$, $E_\gamma=m_\chi$ the gamma-ray line energy and $\rho(r)$ the DM distribution as function of the distance from the GC. The coordinate $s\geq 0$ runs along the line of sight and $r(s,\theta)=\sqrt{r^2_0+s^2-2r_0s\cos\theta}$, where $r_0$ denotes the distance between the Sun and the GC. If DM is not constituted by self-conjugated particles (e.g. Dirac fermions), the equation (\ref{fluxo}) has to be divided by a factor of 2 if only particle-antiparticle annihilations are present.

Weniger (2012) analyzed 43 months of data (from 4 Aug 2008 to 8 Mar 2012) with energies between 1 and 300 GeV. In regions close to the GC, it was found a 4.6$\sigma$ indication for a gamma-ray line at $E_\gamma\approx 130$ GeV. Considering the look-elsewhere effect, the significance of the observed excess is 3.2$\sigma$.  This spectral feature can be interpreted in terms of DM particles annihilating into two photons. The observations imply that $m_\chi=129.8\pm 2.4^{+7}_{-13}$ GeV and $\langle\sigma v\rangle_{\chi\chi\rightarrow\gamma\gamma}=1.27\pm 0.32^{+0.18}_{-0.28}\times 10^{-27} \text{cm}^3\text{s}^{-1}$. It is assumed the Einasto profile for the DM distribution in the Milk Way:
\begin{equation}
\rho_{\text{Ein}}(r)\propto\exp\left[-\frac{2}{\alpha}\left(\frac{r}{r_s}\right)^\alpha\right],
\end{equation}
with $\alpha=0.17$, $r_s=20$ kpc and normalized to $\rho_{\text{Ein}}(r_0)=0.4$ GeV cm$^{-3}$.

\section{The model}

Lee, Kim \& Shin (2008) proposed a renormalizable fermionic dark matter model consisting of three sectors:
\begin{equation}
{\cal L}={\cal L}_{\rm SM}+{\cal L}_{\rm hid}+{\cal L}_{\rm int}.
\end{equation}
The hidden sector is composed of a real scalar field $S$ and a Dirac fermion field $\chi$ wich are Standard Model (SM) gauge singlets. Its Lagrangian is given by
\begin{equation}
{\cal L}_{\rm hid}={\cal L}_{S}+\bar{\chi}\ \left(i\slashed{\partial} -m_{\chi 0}\right)\chi-g_{S}\bar{\chi}\chi S,
\label{hid}
\end{equation}
with
\begin{equation}
{\cal L}_{S}=\frac{1}{2}(\partial_\mu S)(\partial^\mu S)-\frac{m^2_0}{2}S^2-\frac{\lambda_3}{3!}S^3-\frac{\lambda_4}{4!}S^4.
\label{LS}
\end{equation}
The hidden sector and the SM fields interact through
\begin{equation}
{\cal L}_{\rm int}=-\lambda_1 H^{\dag} H S -\lambda_2 H^{\dag} H S^2.
\label{Lint}
\end{equation}
The scalar potential given in (\ref{LS}) and (\ref{Lint}) together with the SM potential $-\mu^{2}H^{\dag}H+\lambda_{0}(H^{\dag}H)^{2}$ lead to the vacuum expectation values (VEVs) $v_0$ for the neutral component of the SM Higgs doublet to give the electroweak symmetry breaking, and $\langle S \rangle=x_0$ for the singlet scalar sector. The extremum conditions $\partial V/\partial H|_{\langle H^{0} \rangle =v_{0}/\sqrt{2}}=0$ and  $\partial V/\partial S|_{\langle S \rangle=x_0}=0$ lead to the relations
\begin{equation}
\begin{split}
\mu^{2} & =\lambda_{0}v_{0}^{2}+\left(\lambda_{1}+\lambda_{2}x_{0}\right)x_{0},\\
m_{0}^{2} & =-\frac{\lambda_{3}}{2}x_{0}-\frac{\lambda_{4}}{6}x_{0}^{2}-\frac{\lambda_{1}v_{0}^{2}}{2x_{0}}-\lambda_{2}v_{0}^{2}.
\end{split}
\end{equation}
The neutral scalar states $h$ and $s$ defined by $H^{0}=(v_{0}+h)/\sqrt{2}$ and $S=x_0+s$ are mixed. The components of the mass matrix are
\begin{equation}
\begin{split}
\mu_{h}^{2} & \equiv\frac{\partial^{2}V}{\partial h^{2}}\bigg|_{h=s=0}=2\lambda_{0}v_{0}^{2}, \\
\mu_{s}^{2} & \equiv\frac{\partial^{2}V}{\partial s^{2}}\bigg|_{h=s=0}=\frac{\lambda_{3}}{2}x_{0}+\frac{\lambda_{4}}{3}x_{0}^{2}-\frac{\lambda_{1}v_{0}^{2}}{2x_{0}}, \\
\mu_{hs}^{2} & \equiv\frac{\partial^{2}V}{\partial h\partial s}\bigg|_{h=s=0}=\left(\lambda_{1}+2\lambda_{2}x_{0}\right)v_{0}.
\end{split}
\end{equation}
The mass eigenstates $h_1$ and $h_2$ are
\begin{equation}
\begin{split}
h_1 & =\sin\theta\, s+\cos\theta\, h, \\
h_2 & =\cos\theta\, s-\sin\theta\, h,
\end{split}
\end{equation}
where the mixing angle is defined by
\begin{equation}
\tan\theta=\frac{y}{1+\sqrt{1+y^{2}}},
\end{equation}
with $y\equiv2\mu_{hs}^{2}/\left(\mu_{h}^{2}-\mu_{s}^{2}\right)$. The mass eigenvalues are given by
\begin{equation}
m_{1,2}^{2}=\frac{\mu_{h}^{2}+\mu_{s}^{2}}{2}\pm\frac{\mu_{h}^{2}-\mu_{s}^{2}}{2}\sqrt{1+y^{2}},
\label{masses}
\end{equation}
where the upper and lower signs correspond respectively to $m_1$ and $m_2$. The definition of $\theta$ ensures that $|\cos\theta|>\frac{1}{\sqrt{2}}$. Therefore $h_1$ is the SM Higgs-like state and $h_2$ is the singlet-like state. 

The singlet fermion mass is $m_\chi=m_{\chi 0}+g_Sx_0$. We take $m_\chi$ as an independent parameter of the model since $m_{\chi 0}$ may be chosen freely.

\section{Relic density}

The thermally averaged annihilation cross section of DM particles and its relic density $\Omega_{\chi}$ are related via
\begin{equation}
\Omega_{\chi}h^{2}\approx\frac{\left(1.07\times10^{9}\text{GeV}^{-1}\right)x_{F}}{\sqrt{g_{*}}M_{P}\langle \sigma_{\text{ann}} v \rangle_F},
\end{equation}
where $g_*$ counts the effective degrees of freedom of the relativistic quantities in equilibrium. The inverse freeze-out temperature $x_F=m_{\chi}/T_F$ is determined by the iterative equation
\begin{equation}
x_{F}=\ln\left(\frac{m}{2\pi^{3}}\sqrt{\frac{45M_{P}^{2}}{2g_{*}x_{F}}}\langle\sigma_{\text{ann}} v\rangle_F\right).
\end{equation}
The thermally averaged cross section is given by
\begin{equation}
\begin{split}
\langle \sigma v \rangle &= 
\frac{1}{8m_{\chi}^{4}TK_{2}^{2}\left(m_{\chi}/T\right)}
\int_{4m_{\chi}^{2}}^{\infty}ds\sigma\left(s\right)\\
&\times\left(s-4m_{\chi}^{2}\right)\sqrt{s}K_{1}\left(\frac{\sqrt{s}}{T}\right),
\end{split}
\end{equation}
where $K_{1,2}$ are the modified Bessel functions (Gondolo \& Gelmini 1991).

We consider $m_\chi=130$ GeV and, typically, the freeze-out temperature is $T_F\simeq 20$. Thus, the dominant final states are $b\bar{b}$, $W^{+}W^{-}$, $ZZ$ and $h_1h_1$. The annihilation of DM particles occur via $h_i$ mediated $s$-channel processes. The total annihilation cross section is given by
\begin{equation}
\begin{split}
\sigma_{\text{ann}}&=\frac{\left(g_{S}\sin\theta \cos\theta\right)^2}{16\pi}\frac{1}{2}\sqrt{1-\frac{4m_{\chi}^{2}}{s}}\\
&\times\xi(s,m_{1},m_{2},\Gamma_{1},\Gamma_{2})\left[\left(\frac{m_{b}}{v_{0}}\right)^{2}6s\left(1-\frac{4m_{b}^{2}}{s}\right)^{3/2}
\right.\\
&\left.+\left(2\frac{m_{W}^{2}}{v_{0}}\right)^{2}\left(2+\frac{\left(s-2m_{W}^{2}\right)^{2}}{4m_{W}^{2}}\right)
\sqrt{1-\frac{4m_{W}^{2}}{s}}\right.\\
&\left.+\frac{1}{2}\left(2\frac{m_{Z}^{2}}{v_{0}}\right)^{2}
\left(2+\frac{\left(s-2m_{Z}^{2}\right)^{2}}{4m_{Z}^{2}}\right)\sqrt{1-\frac{4m_{Z}^{2}}{s}}\right]\\
&+\sigma_{h_1h_1},
\end{split}
\label{crosssection}
\end{equation}
where $\Gamma_{i}$ is the decay width of $h_i$, $\sqrt{s}$ is the center of mass energy and we define
\begin{equation}
\begin{split}
&\xi(s,m_{1},m_{2},\Gamma_{1},\Gamma_{2})\equiv\left[\frac{1}{\left(s-m_{1}^{2}\right)^{2}+m_{1}^{2}\Gamma_{1}^{2}}\right.\\
&\hspace{3.2cm}\left.+\frac{1}{\left(s-m_{2}^{2}\right)^{2}+m_{2}^{2}\Gamma_{2}^{2}}\right.\\
&\left.-\frac{2\left(s-m_{1}^{2}\right)\left(s-m_{2}^{2}\right)
+2m_{1}m_{2}\Gamma_{1}\Gamma_{2}}{\left(\left(s-m_{1}^{2}\right)^{2}
+m_{1}^{2}\Gamma_{1}^{2}\right)\left(\left(s-m_{2}^{2}\right)^{2}+m_{2}^{2}\Gamma_{2}^{2}\right)}\right].
\end{split}
\end{equation}

\section{Annihilation into two photons}

\begin{figure}[t]
\centering
\includegraphics[scale=0.6]{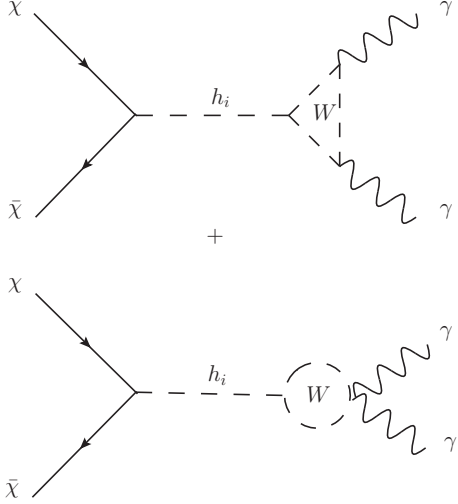}
\caption{Feynman diagrams for the dark matter annihilation into two photons mediated by virtual W bosons.}
\label{diagramas}
\end{figure}

Since dark matter is electrically neutral, it does not directly couple to photons. Annihilation into photons can be generated with loops of charged particles. For the model under consideration, the dominant Feynman diagrams contributing to this process are shown in Fig. \ref{diagramas}.

The gauge invariant $h_i$ decay amplitude into two photons is given by
\begin{equation}
\begin{split}
\mathcal{M}_i&=\frac{e^{2}gM_i}{\left(4\pi\right)^{2}m_{W}}A_1(\beta)\left(k_{1}\cdot k_{2}g^{\mu\nu}-k_{2}^{\mu}k_{1}^{\nu}\right)\\
&\times\varepsilon_{\mu}\left(k1\right)\varepsilon_{\nu}\left(k2\right),
\end{split}
\end{equation}
where
\begin{equation}
A_1(\beta)=2+3\beta+3\beta\left(2-\beta\right)f\left(\beta\right),
\end{equation}
with $\beta=4m_{W}^{2}/m_{i}^{2}$,
\begin{equation}
f\left(\beta\right)=\begin{cases}
\arcsin^{2}\left(\beta^{-\frac{1}{2}}\right)& \text{for $\beta\geq1$}\\
-\frac{1}{4}\left(\ln\frac{1+\sqrt{1-\beta}}{1-\sqrt{1-\beta}}-i\pi\right)^{2}& \text{for $\beta<1$}
\end{cases},
\label{fbeta}
\end{equation}
and
\begin{equation}
M_i=\begin{cases}
\cos\theta & \text{for $i=1$}\\
-\sin\theta& \text{for $i=2$}
\end{cases}.
\end{equation}
In the annihilation process $\chi\bar{\chi}\rightarrow\gamma\gamma$, the $h_i$ particles can be off-shell. Therefore we substitute $m_{i}^{2}$ with $s$ in the definition of $\beta$. The annihilation cross section is given by
\begin{equation}
\begin{split}
\sigma_{\chi\bar{\chi}\rightarrow\gamma\gamma}&=\frac{\left(g_{S}\sin\theta\cos\theta\right)^{2}}{64\pi}\xi(s,m_{1},m_{2},\Gamma_{1},\Gamma_{2})\\
&\times s^{3/2}\sqrt{s-4m_{\chi}^{2}}\left[\frac{e^{2}gF\left(\beta\right)}{\left(4\pi\right)^{2}m_{W}}\right]^2.
\end{split}
\end{equation}

\section{Results and discussions}

The direct annihilation of DM particles into two photons leads to monochromatic gamma rays with $E_{\gamma}=m_{\chi}$, so we take $m_{\chi}=130$ GeV. Moreover, we take $m_{1}=126$ GeV (ATLAS Collaboration 2012) and choose $m_{2}=260.01$ GeV to increase the $\chi\bar{\chi}\rightarrow\gamma\gamma$ cross section. The degrees of freedom parameter is $g_*=86.25$ (Kolb \& Turner 1990). The decay widths were determined by the HDECAY code (Djouadi, Kalinowski \& Spira 1998) with updated data. The other parameters are chosen under the condition (\ref{masses}). 

Demanding the observed relic density, we obtain $g_S=0.655$. Since we consider Dirac fermions, the required annihilation cross section into two photons to produce the observed gamma-ray line is $\langle\sigma v\rangle_{\chi\bar{\chi}\rightarrow\gamma\gamma}=2.54\times 10^{-27} \text{cm}^3\text{s}^{-1}$. The cross section averaged over the DM particles velocities is calculated assuming a Maxwell-Boltzmann distribution with $v_{\text{rms}}=300$ km/s. We obtain $\langle \sigma v \rangle_{\chi\bar{\chi}\rightarrow\gamma\gamma}=1.25\times 10^{-28} \text{cm}^3\text{s}^{-1}$, below than the required.

We propose a modification in the model to enhance $\langle\sigma v\rangle_{\chi\bar{\chi}\rightarrow\gamma\gamma}$. It consists in the introduction of a new scalar multiplet $E$ that carries electric charge and couples to the singlet scalar $S$. Thus, we add the interaction 
\begin{equation}
\mathcal{L}_E=g_ES^2E^{\dag}E
\label{LE}
\end{equation}
to the Lagrangian of the hidden sector. To avoid charged relics, $E$ must be unstable. Also, to enhance the cross section for $\chi\bar{\chi}\rightarrow\gamma\gamma$, we may let $E$ 
transform under QCD or a hidden SU(N) gauge symmetry. 

The gauge invariant $h_i$ decay amplitude into two photons mediated by $E$ is given by
\begin{equation}
\begin{split}
\mathcal{M}^{'}_i&=-\frac{N_c(q_Ee)^{2}g_Ev_0M^{'}_i}{\left(4\pi\right)^{2}m^2_{E}}A_0(\beta)\\
&\times(k_{1}\cdot k_{2}g^{\mu\nu}-k_{2}^{\mu}k_{1}^{\nu})\varepsilon_{\mu}(k_1)\varepsilon_{\nu}(k_2),
\label{Mhgg}
\end{split}
\end{equation}
where 
\begin{equation}
A_0(\beta)=\beta-\beta^2f(\beta),
\end{equation}
with $\beta=4m_{E}^{2}/m_{i}^{2}$, $f(\beta)$ following the definition (\ref{fbeta}), and 
\begin{equation}
M^{'}_i=\begin{cases}
\sin\theta & \text{for $i=1$}\\
\cos\theta& \text{for $i=2$}
\end{cases}.
\end{equation}
The quantities $q_E$ and $N_c$ are, respectively, the electric charge in units of $e$ and the number of colors of QCD or the hidden SU(N) gauge symmetry. The annihilation cross section into two photons is given by
\begin{equation}
\begin{split}
\sigma_{\chi\bar{\chi}\rightarrow\gamma\gamma}&=\frac{g_s^2}{32\pi}s^{3/2}\sqrt{s-4m_{\chi}^{2}}\left[\frac{(q_Ee)^{2}N_cg_Ex_0A_0(\beta)}{\left(4\pi\right)^{2}m^2_E}\right]^2\\
&\times\xi^{'}(s,\theta,m_{1},m_{2},\Gamma_{1},\Gamma_{2}),
\label{sigmaxxgg}
\end{split}
\end{equation}
where we define
\begin{equation}
\begin{split}
&\xi^{'}(s,\theta,m_{1},m_{2},\Gamma_{1},\Gamma_{2})\equiv\left[\frac{\sin^4\theta}{\left(s-m_{1}^{2}\right)^{2}+m_{1}^{2}\Gamma_{1}^{2}}\right.\\
&\left.+\frac{\cos^4\theta}{\left(s-m_{2}^{2}\right)^{2}+m_{2}^{2}\Gamma_{2}^{2}}-\sin^2\theta\cos^2\theta\right.\\
&\left.\times\frac{2\left(s-m_{1}^{2}\right)\left(s-m_{2}^{2}\right)
+2m_{1}m_{2}\Gamma_{1}\Gamma_{2}}{\left(\left(s-m_{1}^{2}\right)^{2}
+m_{1}^{2}\Gamma_{1}^{2}\right)\left(\left(s-m_{2}^{2}\right)^{2}+m_{2}^{2}\Gamma_{2}^{2}\right)}\right].
\end{split}
\end{equation}

The process $\chi\bar{\chi}\rightarrow E\bar{E}$ contributes to the total annihilation cross section with
\begin{equation}
\begin{split}
\sigma_{\chi\bar{\chi}\rightarrow E\bar{E}}&=\frac{(g_Sg_Ex_0)^2N_c}{8\pi}\sqrt{1-\frac{4m_{\chi}^{2}}{s}}\sqrt{1-\frac{4m_{E}^{2}}{s}}\\
&\times\xi^{'}(s,\theta,m_{1},m_{2},\Gamma_{1},\Gamma_{2}).
\end{split}
\label{sigmaxxee}
\end{equation}

\begin{figure}[t]
\centering
\includegraphics[scale=0.36]{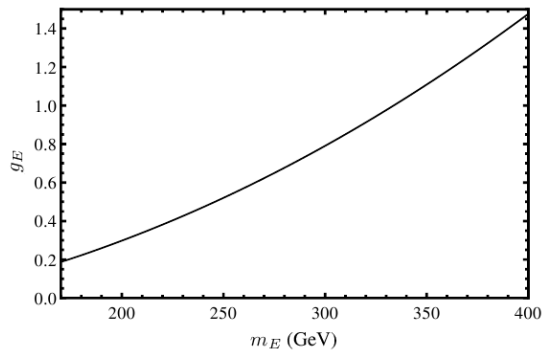}
\caption{Coupling constant $g_E$ as a function of $m_E$, demanding the cross section to produce the Fermi-LAT gamma-ray line, assuming $N_c=1$ and $q_E=1$.}
\label{graph}
\end{figure}

Taking for $\langle \sigma v \rangle_{\chi\bar{\chi}\rightarrow\gamma\gamma}$ the value required to explain the Fermi-LAT gamma-ray line, $N_c=1$ (without enhancement) and $q_E=1$, we find the relation between $g_E$  and $m_E$ shown in Fig. \ref{graph}. We note that the model is under perturbative control up to $m_E\sim 335$ GeV.

\begin{figure}[t]
\centering
\includegraphics[scale=0.35]{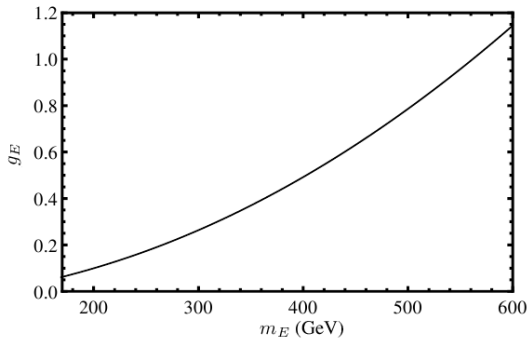}
\caption{Coupling constant $g_E$ as a function of $m_E$, demanding the cross section to produce the Fermi-LAT gamma-ray line, assuming $N_c=3$ and $q_E=1$.}
\label{graph2}
\end{figure}

Assuming $N_c=3$, we find the relation between $g_E$  and $m_E$ shown in Fig. \ref{graph2}. In this case, the model is under perturbative control up to $m_E\sim 560$ GeV. In both situations, with the additional contributions of (\ref{sigmaxxgg}) and (\ref{sigmaxxee}), the coupling constant $g_S$ consistent with the observed relic density is $g_S\approx 0.575$ for the whole range of masses considered. For higher values of $m_E$, one can consider $q_E>1$ to maintain $g_E<1$.

The interaction (\ref{LE}) leads to an enhancement in the Higgs decay into two photons branching ratio. The signal strength parameters are defined as
\begin{equation}
\mu_i=\frac{[\sigma_{j\rightarrow h}\times BR(h\rightarrow i)]_{\text{observed}}}{[\sigma_{j\rightarrow h}\times BR(h\rightarrow i)]_{\text{SM}}},
\end{equation}
for a production of a Higgs that decays into a visible channel $i$ with branching ratio $BR(h\rightarrow i)$. The label $j$ in the cross section denotes that signal events in some final states are defined to be summed over a subset of Higgs production processes $j$. The situation of the measured signal strength of the $h\rightarrow\gamma\gamma$ channel has not been fully elucidated. The ATLAS experiment measured $\mu_{\gamma\gamma}=1.65\pm 0.24$ (Consonni 2013). In contrast, the CMS experiment measured $\mu_{\gamma\gamma}=0.78^{+0.28}_{-0.26}$ (or $\mu_{\gamma\gamma}=1.11^{+0.32}_{-0.30}$ depending on the analysis method), consistent with the Standard Model expectation (Palmer 2013).

We may aporoximate the signal strength of the $h\rightarrow\gamma\gamma$ channel as
\begin{equation}
\mu_{\gamma\gamma}\approx \frac{\Gamma(h\rightarrow\gamma\gamma)^{'}}{\Gamma(h\rightarrow\gamma\gamma)_{SM}},
\end{equation}
where $\Gamma(h\rightarrow\gamma\gamma)^{'}$ is the decay width that takes into account the extra contribution and $\Gamma(h\rightarrow\gamma\gamma)_{SM}$ is the SM decay width.

The SM decay width is given by
\begin{equation}
\begin{split}
\Gamma(h\rightarrow\gamma\gamma)_{SM}&=\frac{G_F\alpha^2m^3_{1}}{128\sqrt{2}\pi^3}\Bigl|A_1(\beta)\\
&+\sum_fN_cq^2_fA_{1/2}(\beta)\Bigr|^2,
\end{split}
\end{equation}
where
\begin{equation}
\begin{split}
A_{1/2}(\beta)=-2\beta[1+(1-\beta)f(\beta)],
\end{split}
\end{equation}
with $\beta=4m^2_f/m^2_1$,  and $f(\beta)$ following the definition (\ref{fbeta}). The quantities $N_c$ and $q_f$ are, respectively, a color factor ($N_c=1$ for leptons and $N_c=3$ for quarks) and the electric charge of the fermion $f$ in units of $e$.

The decay width with the additional contribution from the process mediated by $E$ is
\begin{equation}
\begin{split}
\Gamma(h\rightarrow\gamma\gamma)^{'}&=\frac{G_F\alpha^2m^3_{1}}{128\sqrt{2}\pi^3}\Bigl|A_1(\beta)\\
&+\sum_fN_cq^2_fA_{1/2}(\beta)\\
&-\frac{m_W}{gm^2_E}2g_Ex_0A_0(\beta)\Bigr|^2.
\end{split}
\end{equation}

\begin{figure}[t]
\centering
\includegraphics[scale=0.35]{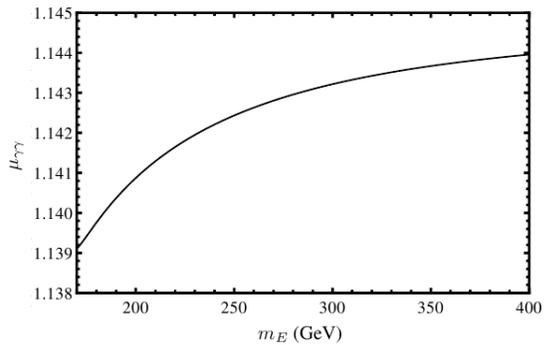}
\caption{Signal strength of the $h\rightarrow\gamma\gamma$ channel as a function of $m_E$ assuming the relation in Fig. \ref{graph}.}
\label{graph3}
\end{figure}

\begin{figure}[t]
\centering
\includegraphics[scale=0.35]{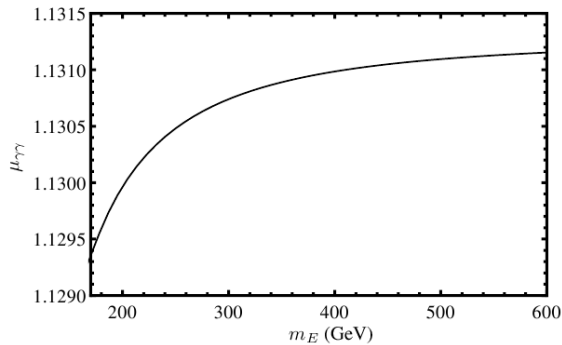}
\caption{Signal strength of the $h\rightarrow\gamma\gamma$ channel as a function of $m_E$ assuming the relation in Fig. \ref{graph2}.}
\label{graph4}
\end{figure}

The signal strength as a function of $m_E$, demanding the Fermi-LAT cross section, is shown in Figs. \ref{graph3} and \ref{graph4} for $N_c=1$ and $N_c=3$, respectively. In both cases, it is consistent with the measurement from the CMS experiment. If confirmed, the ATLAS experiment larger excess is due to another process.

\section{Conclusion}

We have considered a minimal extension of the SM with a singlet Dirac fermion as cold dark matter and a singlet scalar that couples to the Higgs. The annihilation cross section into two photons is smaller than the needed to account for the Fermi-LAT gamma-ray line, so we have added a scalar multiplet carrying electric charge and succeeded in producing the signal. The resulting increase of the decay width for the $h\rightarrow\gamma\gamma$ process is slight and consistent with the measurement from the CMS experiment. The contributions from other annihilations channels such as $\gamma Z$ and $\gamma h_1$ may be addressed in a future work.


\end{document}